\documentclass[twocolumn,trackchanges]{aastex6}
\pdfoutput=1
\usepackage{amsmath}
\usepackage{multirow}

\slugcomment{Published in Astrophysical Journal}
\shorttitle{FRi\reflectbox{3}D: A novel 3D model of CMEs}
\shortauthors{A. Isavnin}

\begin{document}

\newcommand{\insitu}{\textit{in-situ}}
\newcommand{\ie}{\textit{i.e.}}
\newcommand{\eg}{\textit{e.g.}}
\newcommand{\aka}{\textit{a.k.a.}}
\newcommand{\vv}{\textit{vice versa}}

\title{FRi\reflectbox{3}D: A novel three-dimensional model of coronal mass ejections}
\author{A. Isavnin}
\affil{Department of Physics, University of Helsinki, Helsinki, Finland}
\email{alexey.isavnin@helsinki.fi}

\begin{abstract}
We present a novel three-dimensional (3D) model of coronal mass ejections (CMEs) that unifies all key evolutionary aspects of CMEs and encapsulates their 3D magnetic field configuration.
This fully analytic model is capable of reproducing the global geometrical shape of a CME with all major deformations taken into account, \ie{}, deflection, rotation, expansion, "pancaking", front flattening and rotational skew.
Encapsulation of 3D magnetic structure allows the model to reproduce \insitu{} measurements of magnetic field for trajectories of spacecraft--CME encounters of any degree of complexity.
As such, the model can be used single-handedly for consistent analysis of both remote and \insitu{} observations of CMEs at any heliocentric distance.
We demonstrate the latter by successfully applying the model for analysis of two CMEs.
\end{abstract}
\keywords{methods: data analysis --- Sun: coronal mass ejections (CMEs)}

\section{Introduction}
Coronal mass ejections (CMEs) are large-scale explosive eruptions of magnetized plasma from the Sun into the heliosphere.
In addition to being one of the most spectacular manifestations of solar activity these phenomena are the strongest drivers of space weather and one of the major hazards for space exploration \citep{Daglis2001}.
A useful space weather forecast in relation to a CME is expected to predict reliably both time and strength of its impact on space environment.
Both of these characteristics strongly depend on global geometry and internal structure of CME and its evolution \citep{Lee2014}.

Our understanding of CMEs has largely improved over the last two decades due to increasingly detailed remote and \insitu{} observations of the Sun and advances in modeling and simulation techniques.
High-resolution extreme ultraviolet observations have given insight on mechanisms of CME initiation and strong support for their underlying magnetic flux-rope structure \citep{Vourlidas2014}.
Flux-rope eruption is thus the most favorable mechanism of CME known to date \citep{Chen2011}.
Flux-rope formation prior to ejection was further confirmed by extreme ultraviolet observations of the solar disk \citep{Patsourakos2013}.
Stereoscopic white-light coronagraph observations of solar eruptions have given rise to three-dimensional (3D) geometrical modeling of CMEs \citep{Thernisien2006,Thernisien2009} \aka{} forward modeling (FM).
The latter facilitated the studies of CME deflections and rotations \citep{Gui2011,Vourlidas2011} and propagation dynamics \citep{Poomvises2010} in the inner heliosphere.
Heliospheric imaging has provided a way to study CME propagation all the way from Sun to Earth \citep{Eyles2009}.
The emerged modeling techniques facilitated the development of propagation tools for estimation of CME arrival times \citep{Lugaz2009,Davies2013,Moestl2013,Rollett2013}.
Various flux-rope fitting models and reconstruction techniques have been developed, which allow to infer local properties of flux-rope CMEs using single- or multi-spacecraft \insitu{} measurements \citep{Hidalgo2002,Hu2002,Owens2006,Moestl2009,Isavnin2011,NievesChinchilla2016}.

By combining remote-sensing and \insitu{} analysis techniques at different heliocentric distances it has become possible to study evolution of CMEs during their propagation through interplanetary space \citep{Yurchyshyn2009,Isavnin2013,Isavnin2014,Kay2013} and their internal configuration \citep{Kilpua2013}.
However, majority of current techniques and models consider only limited subsets of CME properties and often make inconsistent assumptions about its structure.
Hence, attempts to combine different models to gain the full picture of a CME have limited effectiveness.
Another promising approach to this challenge is empirical 3D modeling of CMEs.
Given the large amount of \insitu{} observations of these structures it has become possible to deduce mean statistical configuration of CMEs outside coronagraph field of view \citep{Janvier2013} as well as of shock wave fronts associated with them \citep{Demoulin2016}.
These studies show that such phenomena as CMEs have certain generic 3D morphology.
Therefore the complexity of each individual CME results from specific deformations that it experienced in the interplanetary medium.

Major evolutionary deformations that CMEs experience during propagation through interplanetary space can be classified into: self-similar expansion, change of orientation (deflections and rotations), front flattening, kinematic distortion due to radial expansion \aka{} "pancaking" and rotational skewing due to rotation of the Sun.
Internal magnetic field structure of a CME also undergoes changes consequent to its deformations.
Impact of a CME on space environment is directly related to its magnetic field configuration at a given location, which in turn depends on its global 3D geometrical and morphological structure.

In this work, we present a novel 3D model of CMEs that is capable of reproducing all of their major deformations.
The model embeds also 3D magnetic field structure and is capable of describing both remote and \insitu{} observations of CMEs.
We demonstrate the performance of the model with two case studies of CMEs.

\section{Model}
\begin{figure}[b]
\gridline{
	\fig{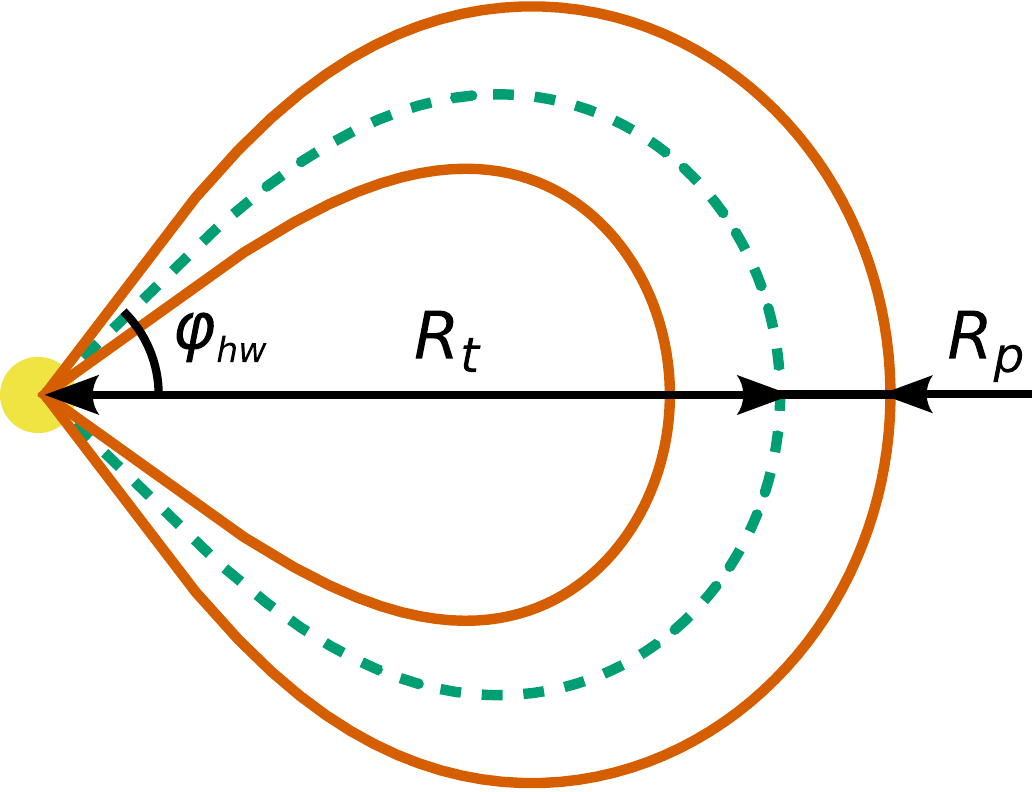}{0.25\textwidth}{(a)}
}
\gridline{
    \fig{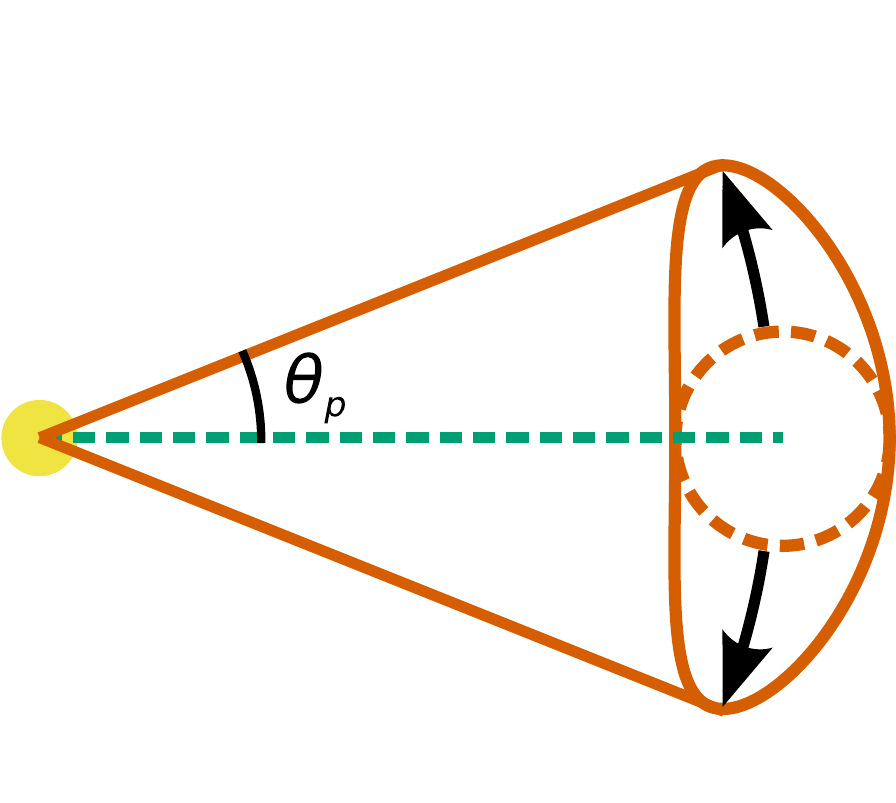}{0.21\textwidth}{(b)}
    \fig{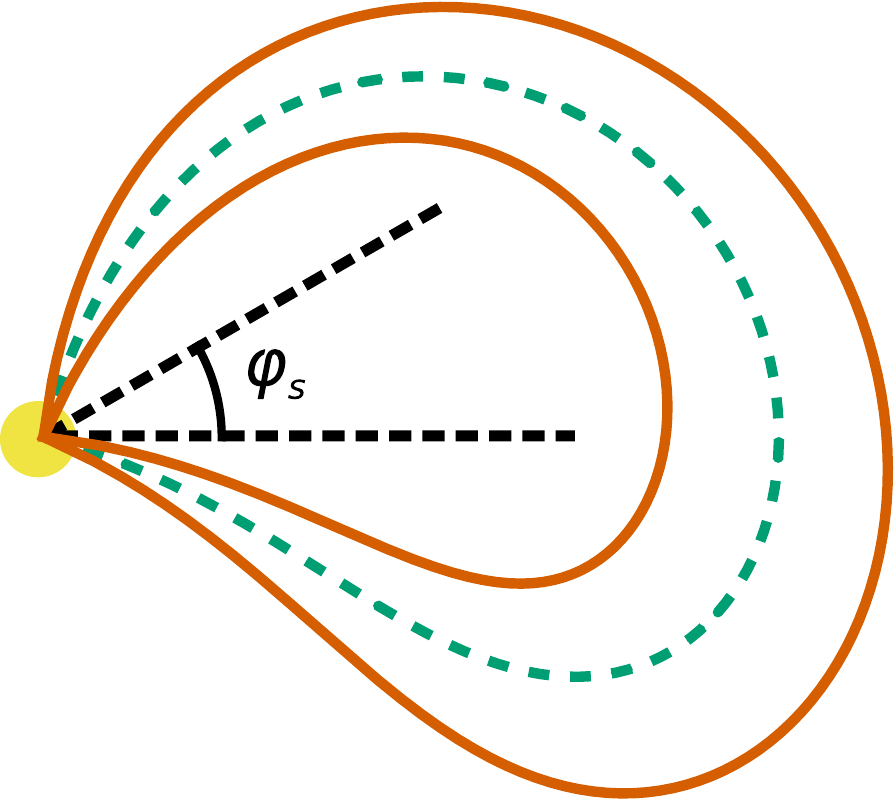}{0.21\textwidth}{(c)}
}
\caption{Schematic representation of a CME (a) and its pancaking (b) and skewing (c) deformations.}
\label{fig:scheme}
\end{figure}
We start by defining a 3D shell of CME and then we populate it with magnetic field.
First, we consider a simplified representation of CME as a bunch of magnetic field lines attached by both ends to the Sun and forming a croissant-like shape (Fig.~\ref{fig:scheme}a).
Heliocentric distance to the apex (the furthermost point) of the axis of the structure is its toroidal height $R_t$.
We further assume that this quasi-CME has circular cross-section anywhere perpendicular to its axis.
The radius of the cross-section varies proportionally to the heliocentric distance with the largest in the apex of the structure (we call it poloidal height and denote as $R_p$) and tending to zero in the Sun as 
\begin{equation}
R(\varphi)=\frac{D(\varphi)}{2}=\frac{R_p}{R_t}r(\varphi),
\label{eq:diameter}
\end{equation}
where $R(\varphi)$ and $D(\varphi)$ are radius and diameter of the cross-section and $r(\varphi)$ describes the axis of the structure in polar coordinates.

\begin{figure}[t]
\gridline{
	\fig{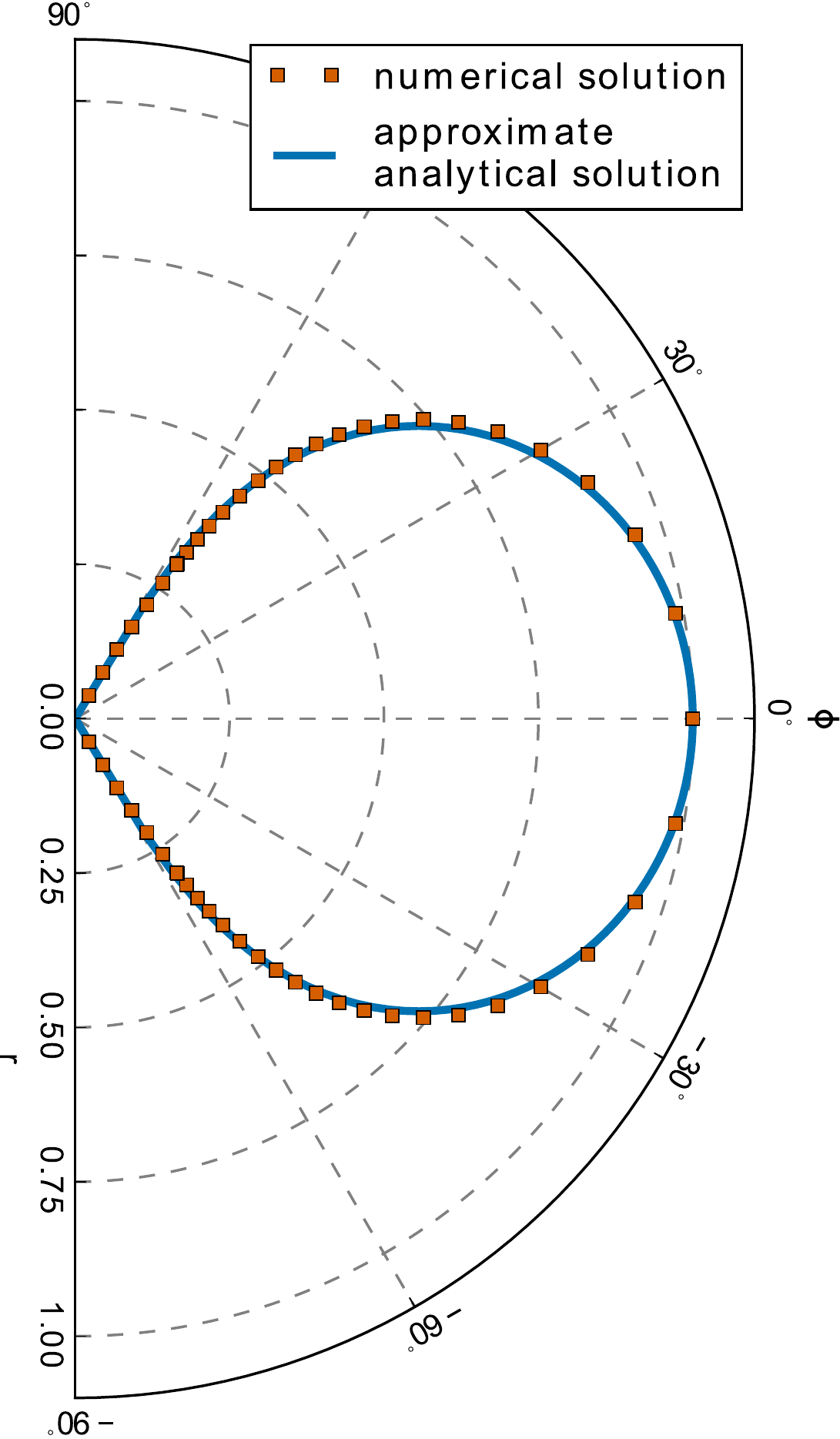}{0.2\textwidth}{(a)}
    \fig{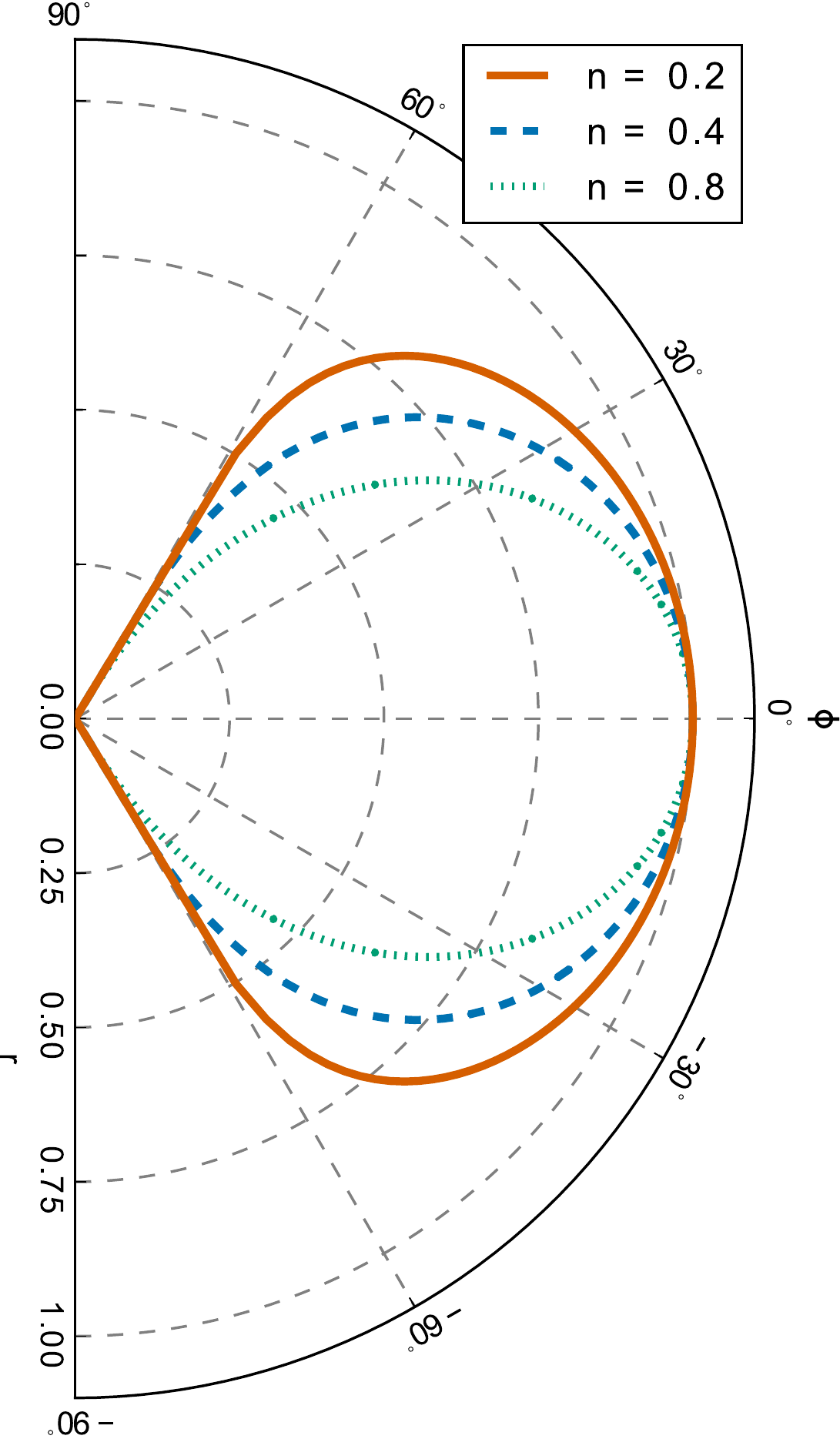}{0.2\textwidth}{(b)}
}
\caption{Numerical and approximate solutions to balance equation (a). Front flattening of the approximate solution (b). The half-width was kept constant at $60^\circ$.}\label{fig:axis}
\end{figure}

This 3D structure is assumed to be in equilibrium in the stream of hydrodynamic wind radially outflowing from the Sun with constant speed.
By equilibrium we understand the balance between the forces of magnetic tension ($\mathbf{F_B}$), gravity ($\mathbf{F_G}$) and hydrodynamic streamlining ($\mathbf{F_H}$):
\begin{equation}
\mathbf{F_H}=\mathbf{F_G}+\mathbf{F_B}
\label{eq:balance}
\end{equation}
For the sake of simplicity we assume the equilibrium to be quasistatic, \ie{}, the structure is similar to a non-propagating slingshot in a radial outflow.
In our simplified description we assume the background solar wind to be purely hydrodynamic and non-magnetized. 
Eq.~\ref{eq:balance} does not take into account magnetic pressure since there is no magnetic interaction between the structure and the background wind.
For a small piece of the structure along its axis the curvature of the axis and the variability of poloidal radius can be neglected thus making it reasonable to use cylindrical coordinates.
The radial and azimuth projections of the balance Eq.~\ref{eq:balance} for a small section of the structure along its axis can then be written as
\begin{equation}
dF_D=dF_G+dF_B\cos{\alpha},
\end{equation}
\begin{equation}
dF_L=dF_B\sin{\alpha},
\label{eq:balance-azimuth}
\end{equation}
where $dF_D$ and $dF_L$ are the forces of hydrodynamic drag and lift that act in radial and azimuth directions respectively and $\alpha(\varphi)$ is the angle between normal to the axis and radial direction determined as
\begin{equation}
\cos^2{\alpha}=\frac{r^2}{r^2+r'^2}
\label{eq:normal-angle}
\end{equation}
Assuming that the shape of the structure can be locally described as a cylinder with diameter given by Eq.~\ref{eq:diameter}, drag and lift forces can be estimated as
\begin{equation}
dF_D=\frac{1}{2}\rho v^2C_D(\alpha)D(\varphi)ds,
\end{equation}
\begin{equation}
dF_L=\frac{1}{2}\rho v^2C_L(\alpha)D(\varphi)ds,
\label{eq:lift-force}
\end{equation}
where $\rho$ and $v$ are the density and velocity of the radial outflow. 
$C_D$ and $C_L$ are the drag and lift coefficients that can be estimated for a cylinder \citep{Vakil2009} as
\begin{equation}
C_D(\alpha)=\frac{C_D^0}{2}(1+\cos{2\alpha}),
\end{equation}
\begin{equation}
C_L(\alpha)=C_L^0\sin{2\alpha}.
\label{eq:lift-coefficient}
\end{equation}
$C_D^0$ is the maximum of the drag coefficient of a cylinder, which happens when it is positioned perpendicularly to the flow.
$C_L^0$ is the maximum of the lift coefficient, which happens when a cylinder is positioned at $\pi/4$ angle to the flow.
Using Eqs. \ref{eq:lift-force} and \ref{eq:lift-coefficient} we can rewrite Eq. \ref{eq:balance-azimuth} as
\begin{equation}
\rho v^2C_L^0D(\varphi)\cos{\alpha}=\frac{B_0^2}{2\mu_0}\kappa(\varphi),
\label{eq:balance-azimuth-expanded}
\end{equation}
where $\kappa(\varphi)$ is the curvature of the axis of the structure defined in polar coordinates as
\begin{equation}
\kappa(\phi)=\frac{1}{R_c(\phi)}=\frac{r^2+2r'^2-rr''}{(r^2+r'^2)^{3/2}},
\label{eq:curvature}
\end{equation}
where $R_c(\varphi)$ is the curvature radius.
Putting together Eqs. \ref{eq:normal-angle}, \ref{eq:balance-azimuth-expanded} and \ref{eq:curvature} we arrive to the following equation:
\begin{equation}
r^2(r^2+2r'^2-rr'')=A(r^2+r'^2),
\label{eq:balance-azimuth-final}
\end{equation}
where $A$ combines all the constants of this simplified problem:
\begin{equation}
A=\frac{\rho v^2 C_L^0 R_p \mu_0}{B_0 R_t}.
\end{equation}
The numerical solution, which describes the axis of the structure, is shown in Fig.~\ref{fig:axis}a along with the guessed approximate solution:
\begin{equation}
r(\varphi)=R_t\cos^n(a\varphi).
\label{eq:axis}
\end{equation}
Here, $a=(\pi/2)/\varphi_{hw}$, where $\varphi_{hw}$ is angular half-width of the axis of the structure.
It can be seen from Fig.~\ref{fig:axis}a that the approximate solution (Eq.~\ref{eq:axis}) to the balance equation describes the axis of the structure exceptionally well compared to the numerical one and thus sufficiently reproduces the physics of our simplified problem.
Hence, we will utilize it in our model for simplicity of further calculations.
Coefficient $n$ regulates the front flattening of the structure (Fig.~\ref{fig:axis}b).
Summarizing, the 3D geometry of our model CME at this stage represents a loop structure attached to the Sun by both ends with axis given by Eq.~\ref{eq:axis} and circular cross-section diameter given by Eq.~\ref{eq:diameter}.

\begin{figure}[b]
\gridline{
	\fig{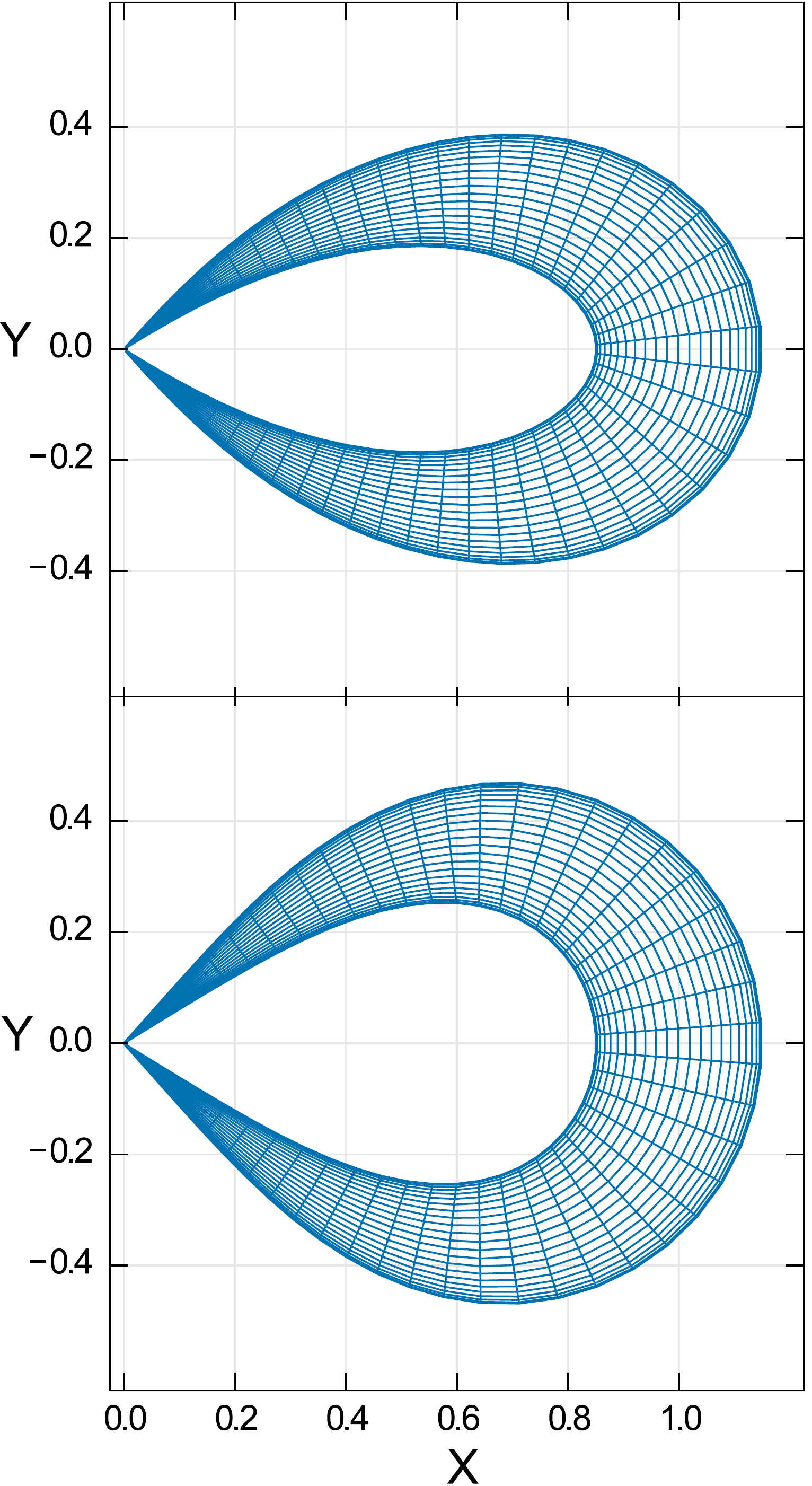}{0.150105\textwidth}{(a)}
    \fig{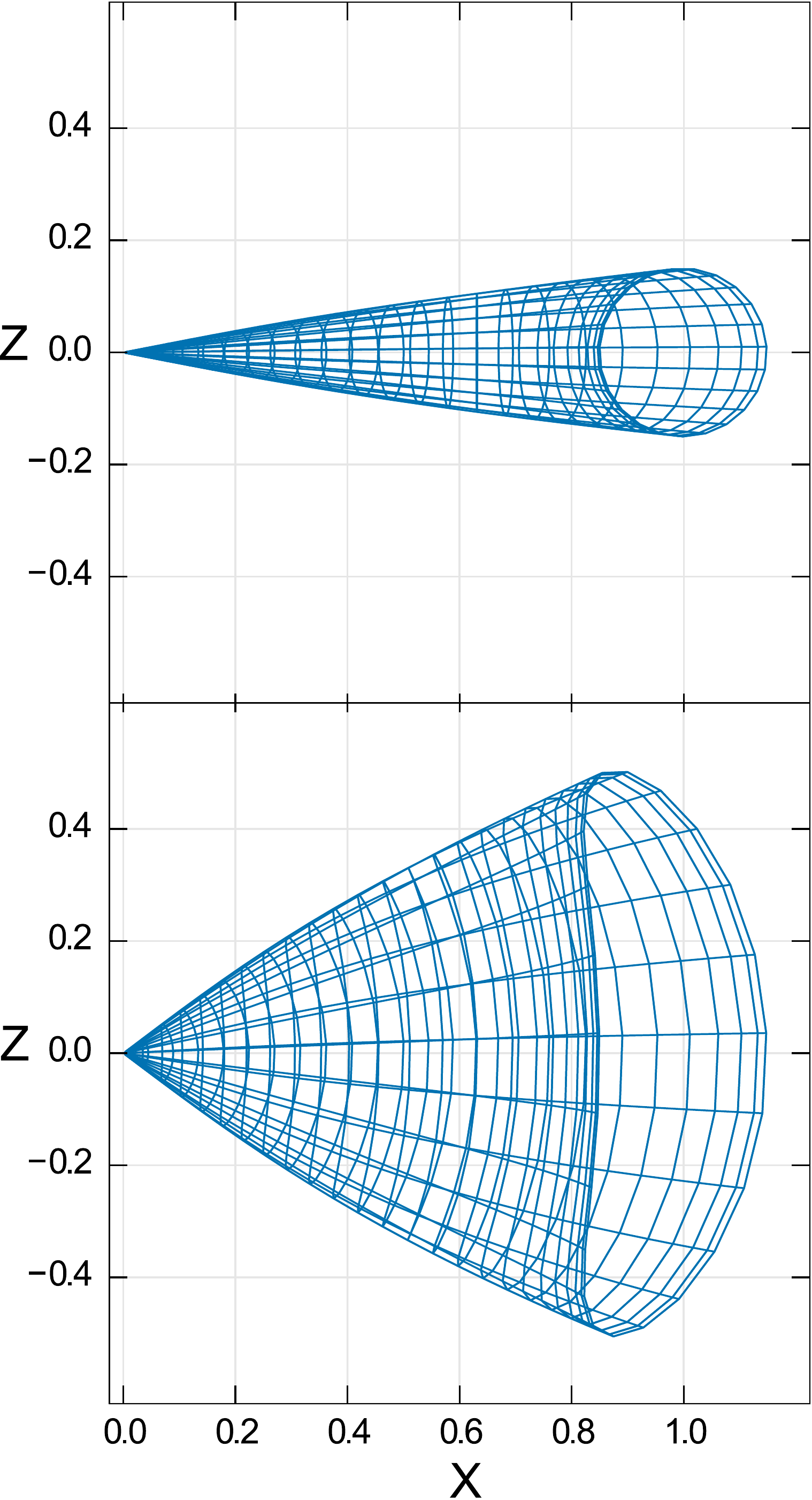}{0.15\textwidth}{(b)}
    \fig{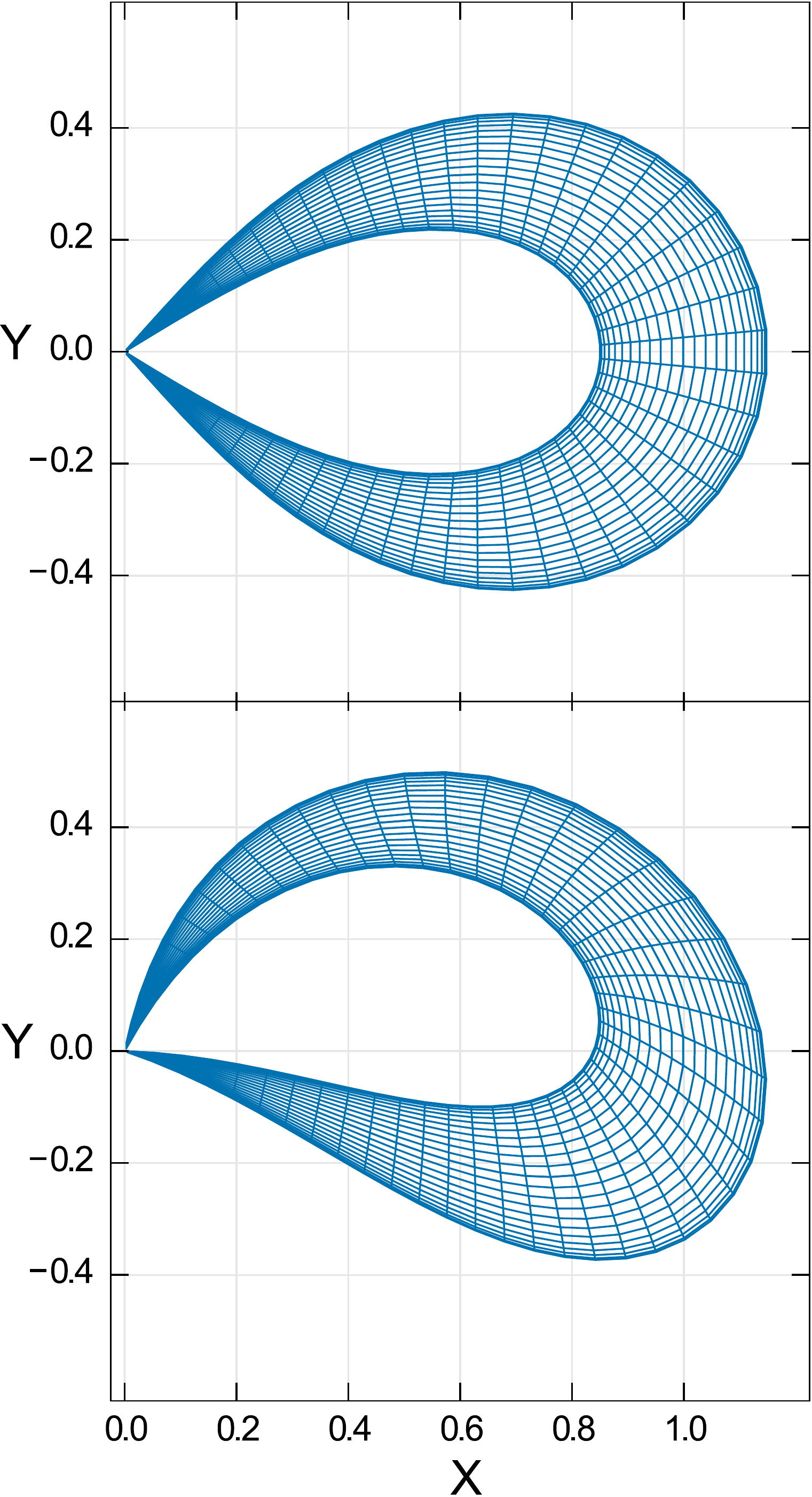}{0.1505\textwidth}{(c)}
}
\caption{Global deformation of 3D CME shell: front flattening (a), pancaking (b) and skewing (c).}
\label{fig:shell}
\end{figure}
Analytic representation of our 3D shell makes it straightforward to apply global deformations to it.
Front flattening deformation happens to CMEs propagating much faster than the speed of the background solar wind \citep{Vrsnak2013}.
The speed difference causes the drag which flattens the front of the CME and slows down its propagation.
This type of deformation is naturally supported by the model through coefficient $n$ (Fig.~\ref{fig:axis}b).
Another type of global deformation often omitted in CME analysis is "pancaking" distortion.
This deformation is a direct consequence of radial propagation of CME through interplanetary space and has purely kinematic nature \citep{Cargill2004,Riley2004}.
We implement this effect in our shell model as a latitudinal stretch, which is characterized by pancaking angle $\theta_p$ (Fig.~\ref{fig:scheme}b).
This parameter describes vertical half-width of a CME and can be considered as a natural counterpart to the lateral half-width $\varphi_{hw}$.
If both half-width $\varphi_{hw}$ and pancaking angle $\theta_p$ do not change during the propagation of a CME one can conclude that its angular size is conserved.
Finally, we implement skew as a rotational deformation around Z axis with the skewing angle $\varphi_s$ (Fig.~\ref{fig:scheme}c).
Skewing happens due to rotation of the Sun and is more pronounced for slow CMEs.
In Fig.~\ref{fig:shell} we demonstrate these three types of global deformations in 3D.
Apparently, the global orientation of the 3D shell is also easily adjustable, which makes it capable of reproducing deflections and rotations.

At this point, our CME shell model is characterized by 9 free parameters: toroidal height $R_t$, poloidal height $R_p$, angular half-width $\varphi_{hw}$, front flattening coefficient $n$, pancaking angle $\theta_p$, skewing angle $\varphi_s$, direction of propagation (latitude $\theta$ and longitude $\varphi$) and tilt angle $\gamma$.

\begin{figure*}[t]
\plotone{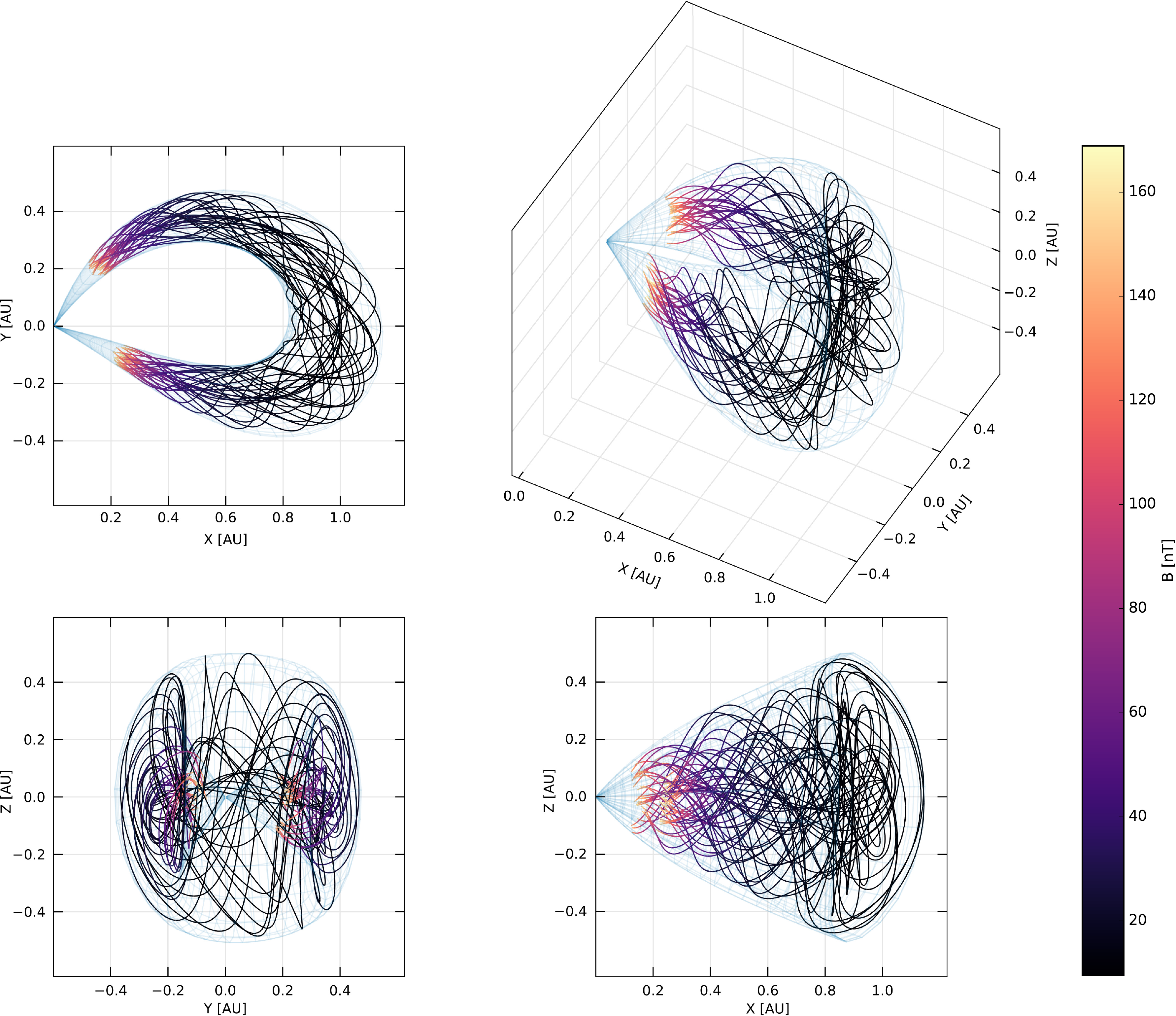}
\caption{FRi3D model of a CME depicted in top, front, side and isometric views. The shell of the model is shown with transparent blue wireframe. Each of 30 randomly selected lines represents an individual magnetic field line. The strength of magnetic field along each line is color-coded. \textit{Note: magnetic lines are not shown close to the Sun, because strong gradient of magnetic field would render color-coding useless.}}
\label{fig:FRi3D}
\end{figure*}
Now that we have a highly flexible 3D shell of a CME we need to populate it with magnetic field.
The inner morphology of a CME is typically described by magnetic flux-rope structure.
Classical flux-rope represents an idealized configuration of magnetic field characterized by the following properties: local cylindrical geometry; helical magnetic field lines with zero twist in the core and increasing to infinity close to the edge of a flux-rope; maximum magnetic field strength along the axis of the flux-rope \citep{Russell1999}.
Such a configuration is often estimated with the Lundquist model, which describes cylindrical magnetic geometry in force-free field \citep{Lundquist1950}.
However, recent studies of field line twist and length distributions within magnetic flux-ropes in CMEs report inconsistencies with the Lundquist model.
\citet{Hu2015} showed that \insitu{} measurements of interplanetary CMEs are consistent with a flux-rope structure with spiral field lines of constant and low twist.
We use the latter finding for construction of the 3D configuration of magnetic field lines for our model.

We start with a collection of parallel magnetic field lines contained in a cylinder of unit radius.
The direction of the magnetic field is characterized by polarity equal to either $+1$ or $-1$, which corresponds to East--West or West--East direction of core magnetic field of a flux-rope.
The length of the cylinder is set to the length $L$ of the axis of the CME shell:
\begin{equation}
L=\int\limits_{-\varphi_{hw}}^{\varphi_{hw}} \left[r^2+\left(\frac{dr}{d\varphi}\right)^2\right]^{1/2} d\varphi.
\end{equation}
The strength of magnetic field is estimated using the distribution of magnetic field from the Lundquist model:
\begin{equation}
B(\rho)=B_0\left[J_0^2(\alpha\rho)+J_1^2(\alpha\rho)\right]^{1/2},
\label{eq:lundquist-distribution}
\end{equation}
where $\rho$ is poloidal distance from the axis, $B_0$ is the strength of the core field, $J_0$ and $J_1$ are Bessel functions of the first and second order and $\alpha\rho$ gives the first zero of $J_0$ at the edge of the flux-rope.
We then apply twisting deformation with constant twist $\tau$, tapering deformation according to Eq.~\ref{eq:diameter} and bend the structure to the shape defined by Eq.~\ref{eq:axis}.
The direction of the twist is characterized by chirality that can be equal to $+1$ or $-1$, which corresponds to right- or left-handedness of a flux-rope respectively.
Thereafter, pancaking and skewing deformations can be applied to the resultant magnetic field structure in the same way as we applied them to the shell earlier.

Lastly, we introduce conservation of magnetic flux $\Phi$ into our model:
\begin{equation}
\Phi=\iint\limits_S \mathbf{B}\cdot\mathbf{ds}.
\label{eq:flux}
\end{equation}
In the simplest case without pancaking and skewing deformations the cross-section of the structure perpendicular to its axis remains circular.
Eq.~\ref{eq:flux} can then be simplified to the following form:
\begin{equation}
\Phi=2\pi\int\limits_0^{\rho_0}B(\rho)\cos(\delta)\rho d\rho,
\label{eq:flux-expanded}
\end{equation}
where
\begin{equation}
\delta=\arctan\frac{2\pi\rho\tau}{L}.
\end{equation}
The flux conservation is introduced by varying axial field $B_0$ in Eq.~\ref{eq:lundquist-distribution} along the axis of the structure so that integral Eq.~\ref{eq:flux-expanded} remains constant, \ie{}, $B_0$ is weakest in the apex and strongest in the footpoints.
After introduction of pancaking and skewing deformations the distribution given by Eq.~\ref{eq:lundquist-distribution} distorts accordingly. 
In such a case in the current version of the model we estimated magnetic flux given by Eq.~\ref{eq:flux} numerically.

\begin{figure}[t]
\plotone{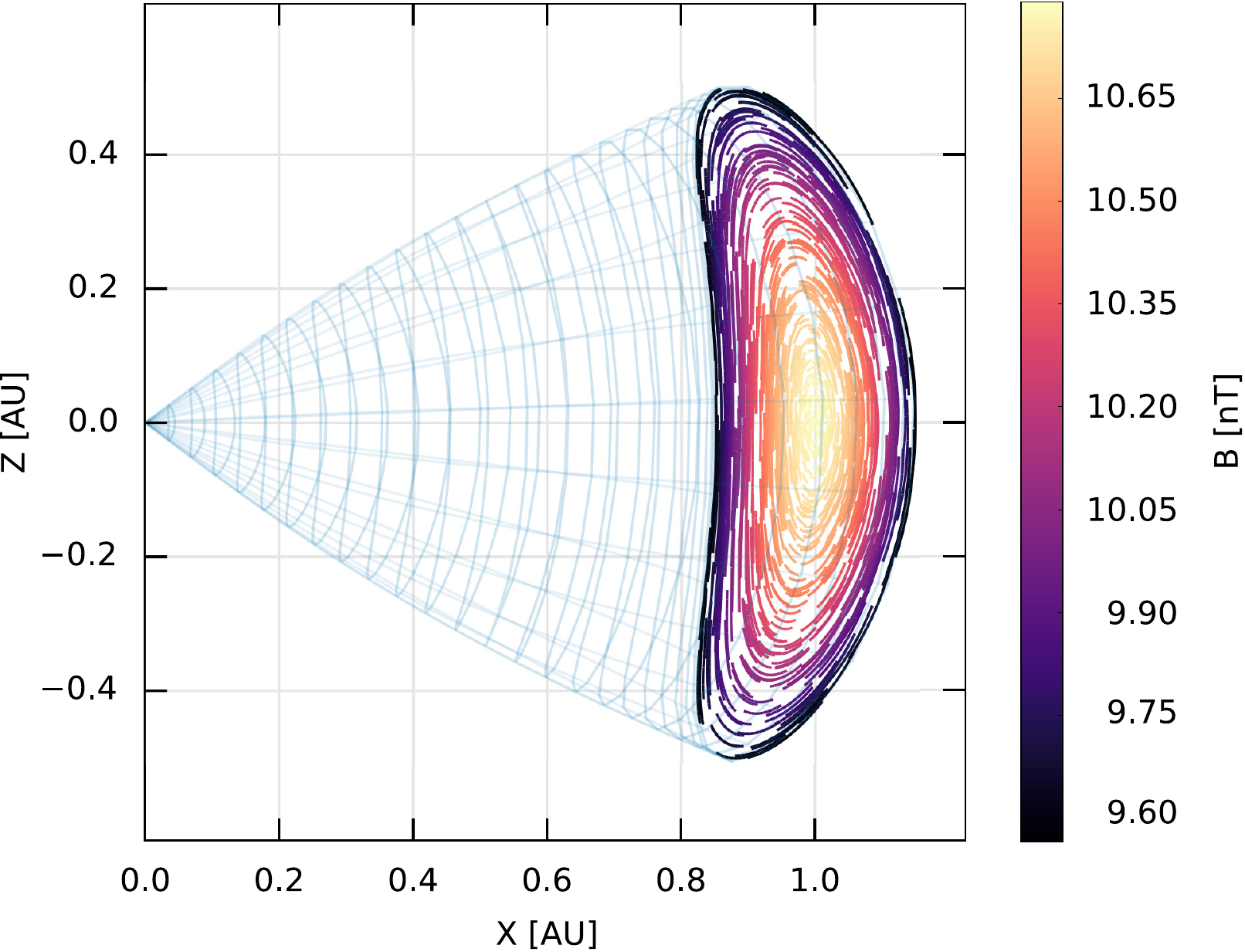}
\caption{Cross-section of the FRi3D model near its apex. The colored lines are the sections of magnetic field lines near the apex. The strength of magnetic field is color-coded.}
\label{fig:FRi3D-slice}
\end{figure}
By adding magnetic structure to the model we introduced two free parameters, \ie{}, twist $\tau$ and magnetic flux $\Phi$, and two binary parameters, \ie{}, polarity and chirality.
The final model of a \textbf{F}lux \textbf{R}ope \textbf{i}n \textbf{3D} (FRi3D) is shown in Fig.~\ref{fig:FRi3D}.
Due to flux conservation the model naturally supports magnetic field expansion.
Fig.~\ref{fig:FRi3D-slice} shows a narrow slice of magnetic field lines of the FRi3D model near its apex.
The shape of the cross-section demonstrates pancaking deformation, while the distribution of magnetic field follows the deformation accordingly.

\begin{figure}[t]
\plotone{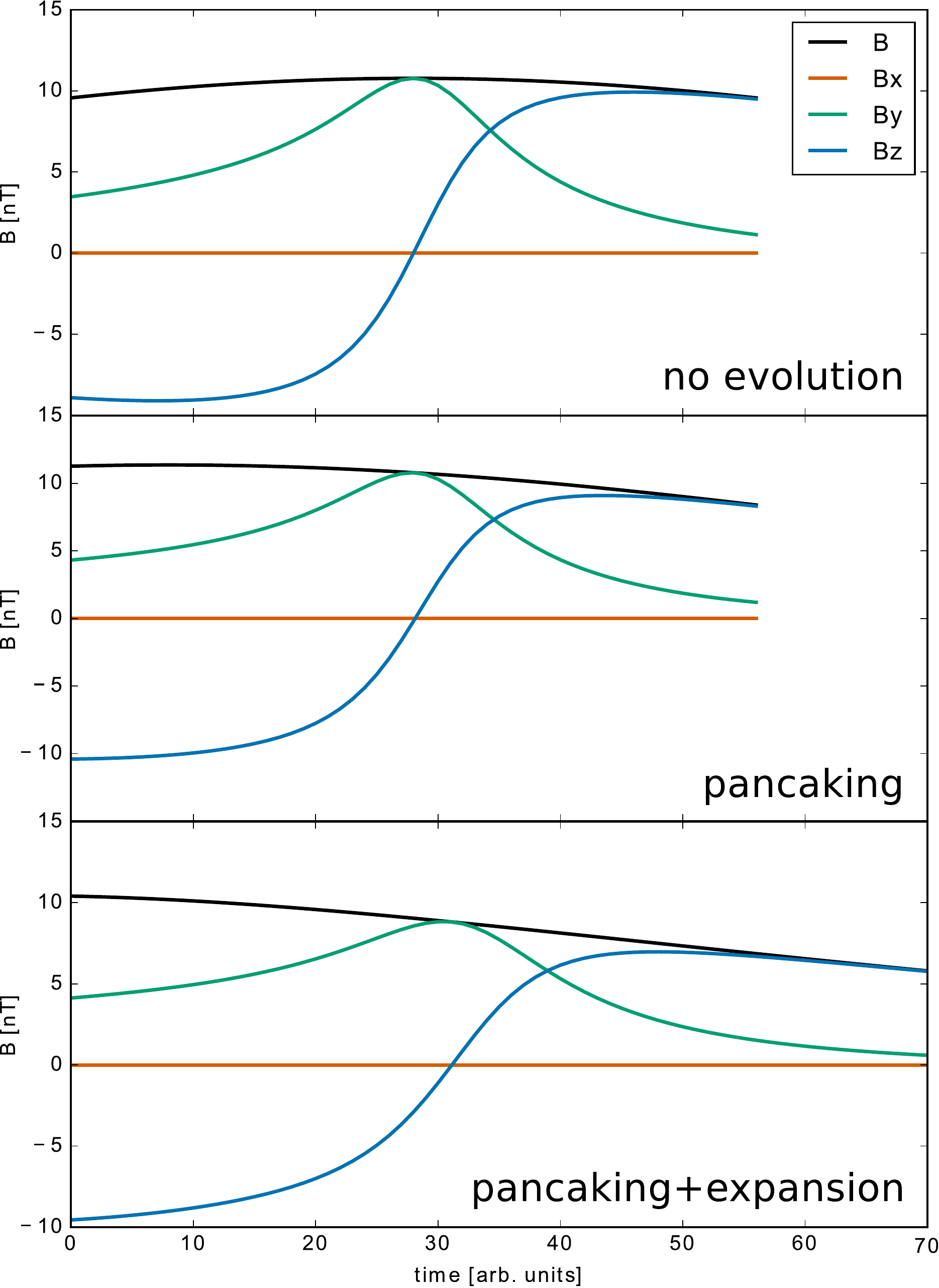}
\caption{Examples of synthetic \insitu{} measurements of magnetic field using the FRi3D model. The top panel shows magnetic field snapshot of a non-evolving CME, the middle panel portrays CME propagation with a fixed pancaking angle $\theta_p$, the bottom panel shows the measurements that take into account expansion (increasing poloidal height $R_p$).}
\label{fig:FRi3D-insitu}
\end{figure}
Simulation of \insitu{} measurements of evolving CMEs is made straightforward and natural with the FRi3D model.
Fig.~\ref{fig:FRi3D-insitu} shows simple examples of such synthetic measurements of magnetic field given in Heliocentric Earth Equatorial coordinate system (HEEQ, \citet{Thompson2006}).
In these examples, it is assumed that a CME is propagating along the Sun--Earth line with zero tilt and is measured \insitu{} by a synthetic spacecraft located in the Lagrangian point 1 (L1).
The model CME has the following parameters: $\theta=0^\circ$, $\varphi=0^\circ$, $\gamma=0^\circ$, $R_p=0.15$~AU, $n=0.5$, $\varphi_{hw}=40^\circ$, $\theta_p=30^\circ$, $\varphi_s=0^\circ$, $\tau=3$, $\Phi=5\mathrm{e}{14}$~Wb, positive polarity and positive chirality.
The top panel of Fig.~\ref{fig:FRi3D-insitu} shows synthetic spacecraft measurements of a non-evolving CME, \ie{}, a snapshot of a magnetic field profile.
One could think of it as a measurement made by a spacecraft passing through a static CME with $R_t=1$~AU and not \vv{}.
The observed rotation of magnetic field is typical for a flux-rope CME.
However, even in such a simplified scenario differences from cylindrical flux-rope models arise.
For example, the asymmetry in $B_y$ component of magnetic field is caused by two 3D geometrical factors: firstly, the bending of magnetic field into a CME shape distorts field lines slightly stronger on the front part of a CME than on a back one; and, secondly, pancaking deformation also distorts magnetic field lines stronger on the front part of a CME than on a back one.
The middle panel of Fig.~\ref{fig:FRi3D-insitu} presents the case of the simplest evolution of a CME.
The CME is set to propagate radially from the Sun by increasing $R_t$, while all other parameters are kept constant:
\begin{equation}
R_t=R_{t0}+V_{R_t}t,
\label{eq:toroidal-height-profile-linear}
\end{equation}
where $V_{R_t}$ is the speed of propagation, \ie{}, the speed of toroidal height growth.
Constancy of $\theta_p$ naturally introduces dynamic pancaking deformation, which gradually increases the area of CME cross-section while it propagates and thus causes magnetic expansion.
That is why the most obvious difference from the previous example is the shift of the maximum of total magnetic field to the start of the measurement.
The total duration of the measurement remained the same, since the CME did not experience any expansion in $R_p$.
Finally, the bottom panel of Fig.~\ref{fig:FRi3D-insitu} shows the same example with added poloidal expansion introduced via increasing $R_p$:
\begin{equation}
R_p=R_{p0}+V_{R_p}t,
\label{eq:poloidal-height-profile-linear}
\end{equation}
where $V_{R_p}$ is the speed of poloidal height growth.
In this case, the total duration of the measurement stretched due to increased radial size of the CME cross-section.
The maximum of total magnetic field shifted to the start of the measurement even more, because magnetic expansion happened at a higher rate.

\section{Case studies}
In this section, we present example case studies of two CMEs using the FRi3D model.
Since we analyze remote and \insitu{} observations with the same model we require both of these measurements to be clear and non-ambiguous.
Both CMEs for our analysis were selected using \textbf{HEL}iospheric \textbf{C}ataloguing \textbf{A}nd \textbf{T}echniques \textbf{S}ervice (HELCATS).

\begin{figure}[t]
\plotone{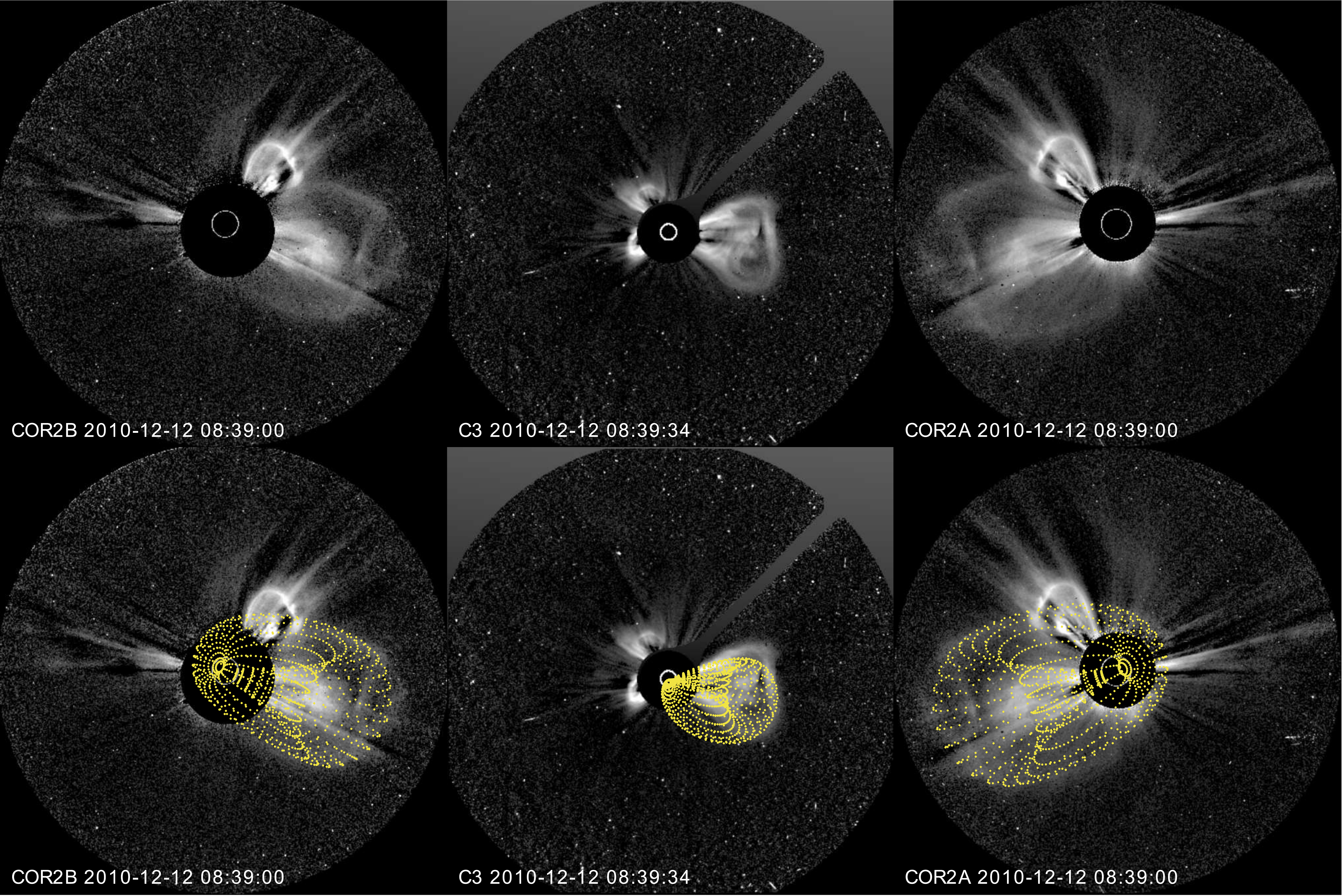}
\caption{Coronagraph images of the CME released on 12 December 2010. The images from left to right show observations from COR2 coronagraph of STEREO-B, C3 coronagraph of SOHO and COR2 coronagraph of STEREO-B respectively. Lower panels show the fitting of the FRi3D model to coronagraph observations.}
\label{fig:20101212-remote}
\end{figure}

In our first case study we present a CME that was released on 12 December 2010 at 02:48~UT as a prominence eruption from the southern hemisphere.
This CME was observed in white-light by coronagraphs onboard SOHO \citep{Domingo1995} and STEREO \citep{Kaiser2008} spacecraft (Fig.~\ref{fig:20101212-remote}).
The erupting loop appeared as a partial halo in STEREO-A field of view and backside partial halo in STEREO-B field of view.
The CME did not produce any visible signatures of a shock wave.
According to SOHO observations the erupting structure smoothly and quickly accelerated to projected speed of 545~km/s and kept it steady during further propagation in coronagraph field of view.

Lower panels of Fig.~\ref{fig:20101212-remote} show the fit of the FRi3D model to coronagraph images from COR2 and C3 instruments onboard STEREO and SOHO spacecraft respectively.
The model performs in a similar way as Graduated Cone Shell model (GCS, \citet{Thernisien2009}) successfully reproducing the the bright flux-rope loop of the CME.
The parameters of the fit are summarized in Table~\ref{tbl:fit-params-201012}.

\begin{figure}[b]
\plotone{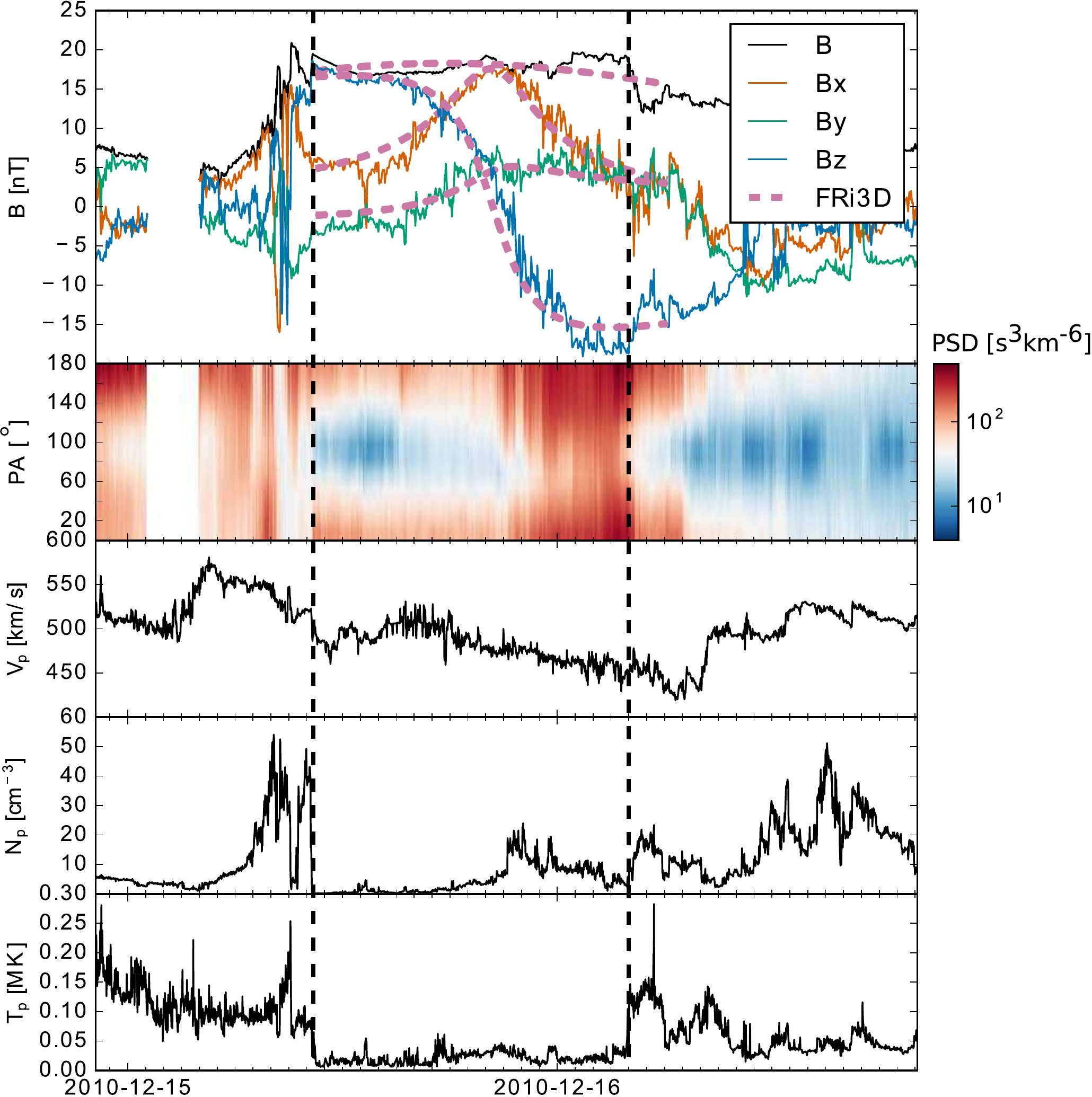}
\caption{\textit{In-situ} magnetic field and plasma measurements of the ICME launched on 12 December 2010 obtained by STEREO-A spacecraft. The panels from top to bottom show magnetic field, electron pitch angle distribution, plasma bulk speed, proton density and proton temperature. Magnetic field data are presented in HEEQ coordinates. Black vertical dash lines show the time range of magnetic obstacle. Purple dash curves show the FRi3D model fit.}
\label{fig:20101215-insitu}
\end{figure}

Interplanetary counterpart of this CME reached STEREO-A spacecraft on 15 December 2010.
The corresponding magnetic obstacle measured between 10:20~UT of 15 December and 04:00~UT of 16 December demonstrated smooth rotation of magnetic field, low proton temperature and proton density as well as bi-directional electron flows (Fig.~\ref{fig:20101215-insitu}), \ie{}, the typical signatures of a magnetic cloud \citep{Zurbuchen2006}.

For the sake of simplicity when fitting the FRi3D model to \insitu{} data we assume that
\begin{itemize}
\item the CME does not experience any evolution apart from pancaking deformation while passing the spacecraft, \ie{}, the case shown in the middle panel of Fig.~\ref{fig:FRi3D-insitu};
\item the speed of CME propagation $V_{R_t}$ is constant and is equal to average speed of magnetic obstacle measured \insitu{};
\item there are no constraints on the geometrical parameters of the CME resulting from our fit to remote data (Fig.~\ref{fig:20101212-remote}), \ie{}, we obtain all 3D geometrical parameters of the CME from \insitu{} data independently from remote observations.
\end{itemize}
The numerical fitting is carried using differential evolution algorithm \citep{Storn1997}, which represents a real-valued version of genetic algorithm.
Differential evolution does not use gradient methods to find the best fit and can search large areas of candidate parameter space.
It does not rely on starting parameters either.
Initial population of possible solutions is chosen randomly from the parameter space.
At each pass through the population the algorithm mutates each candidate solution by mixing with other candidate solutions to create a trial candidate.
The operation continues until sufficiently fit candidate solution is obtained.
The time range of magnetic obstacle is set to be soft, \ie{}, the fitting algorithm is allowed to go beyond the specified temporal boundaries by $\pm2$~hours.
The quality of the fit is assessed by average euclidean distance between the real and synthetic measurements.
The fitting procedure was run several times to ensure the uniqueness of its convergence.
The best fit is shown in Fig.~\ref{fig:20101215-insitu} while the fitted parameters of the FRi3D model are listed in Table~\ref{tbl:fit-params-201012}.
The average euclidean distance between the modeled and real data is 2.85~nT.

\begin{deluxetable}{lrrrrrrr}
\tablecolumns{8}
\tablecaption{Parameters of the FRi3D model fits to remote and \insitu{} data for CME launched on 12 December 2010.\label{tbl:fit-params-201012}}
\tablehead{
\colhead{} & \colhead{$\theta$} & \colhead{$\varphi$} & \colhead{$R_p/R_t$} & \colhead{$\varphi_{hw}$} & \colhead{$\gamma$} & \colhead{$n$} & \colhead{$\theta_p$}
}
\startdata
remote & -14.5 & 55.0 & 0.28 & 55.0 & 16.0 & 0.60 & 23.0\\
\hline
\insitu{} & 0.0 & 59.7 & 0.10 & 66.8 & 0.2 & 0.62 & 29.3\\
\nodata & \multicolumn{7}{l}{$\tau=4.2$, $\Phi=4.7\times10^{14}$~Wb}\\
\nodata & \multicolumn{7}{l}{West--East polarity, right-handed}
\enddata
\end{deluxetable}
Independent fits of the FRi3D model to remote and \insitu{} data show that the strongest geometrical changes were seen in latitude $\theta$ and tilt $\gamma$ of the CME.
According to modeling results it experienced latitudinal deflection and rotation and ended up lying almost perfectly in solar equatorial plane, which agrees with earlier findings by \citet{Isavnin2013,Isavnin2014}.
The analyzed CME experienced overexpansion both in lateral and vertical directions, \ie{}, the increase of half-width $\varphi_{hw}$ and pancaking angle $\theta_p$ \citep{Patsourakos2010}.

\begin{figure*}[t]
\plotone{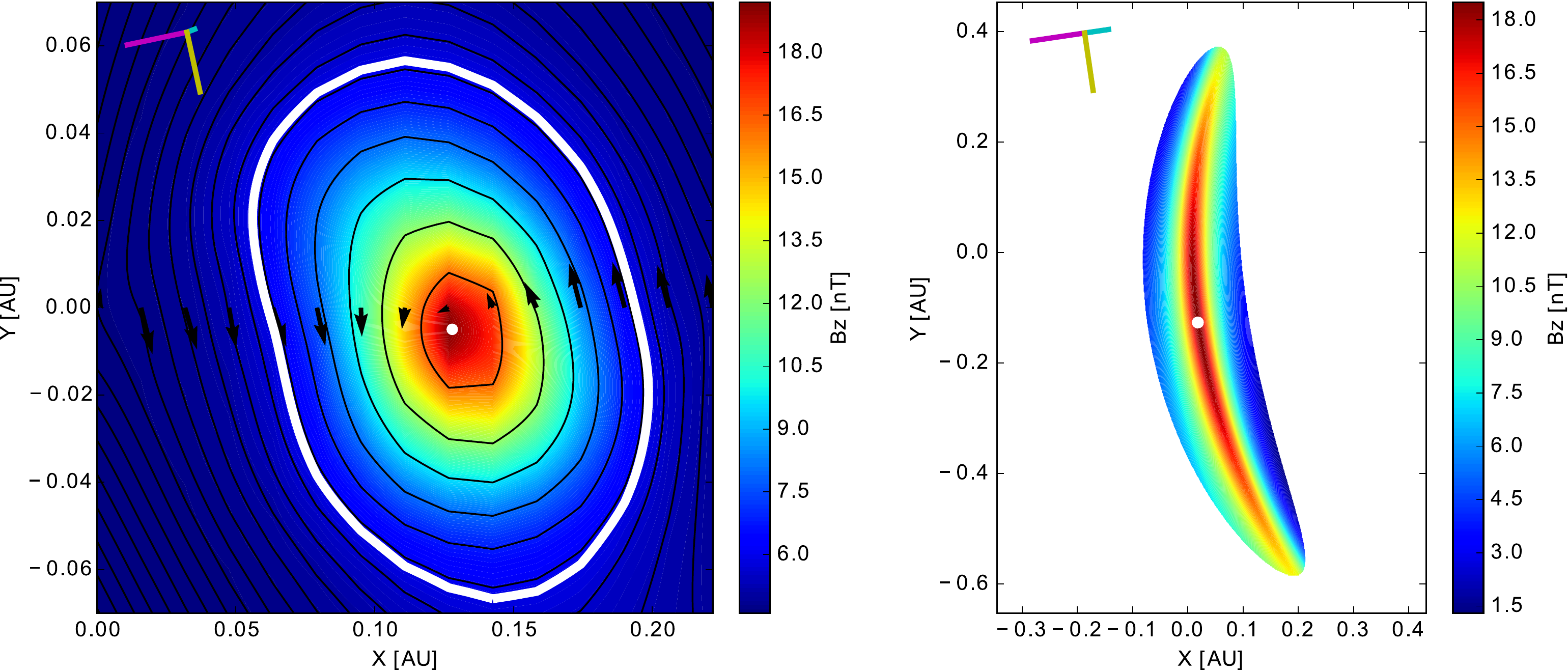}
\caption{Magnetic field maps obtained using Grad-Shafranov reconstruction (left) and FRi3D fitting (right) for the CME released on 12 December 2010. Magnetic field component parallel to the local axis orientation of the CME is color-coded. Local CME axis is marked with the white dot. Projected trajectory of the spacecraft goes along $Y=0$ line. The Sun is to the right. In the top left corners of the maps the projection of HEEQ coordinate system is shown as $X_{HEEQ}$ (cyan), $Y_{HEEQ}$ (magenta) and $Z_{HEEQ}$ (yellow). White solid curve in GS magnetic field map marks the boundary of an unperturbed part of the flux-rope. Black arrows show the projection of magnetic field measurements onto the cross-section plane.}
\label{fig:20101215-cs}
\end{figure*}
We compare our modeling results with two of the most widely used conventional tools for CME research, \ie{}, GCS modeling of remote stereoscopic observations and Grad-Shafranov (GS) reconstruction of \insitu{} measurements.
The results of GCS modeling (not shown in this study) are hard to visually distinguish from the ones of FRi3D modeling.
Indeed on early stages of CME evolution both pancaking and skewing deformations are not pronounced strong enough to demonstrate the discrepancies of the models in a clear way.
Consequently the differences in fitting parameters of the two models lie within the typical error boundaries of forward modeling with the exception of the half-width $\varphi_{hw}$.
The legs of a CME modeled by GCS represent two cones with radially oriented axes while the legs of a FRi3D CME are curved according to Eq.~\ref{eq:axis}.
Therefore the half-width of a FRi3D fit is generally larger than the respective parameter of GCS model even for visually similar fits.

Fig.~\ref{fig:20101215-cs} shows how Grad-Shafranov reconstruction of the analyzed CME compares to corresponding cross-section of the FRi3D model.
Note that the FRi3D model reproduces an evolving non-static CME and hence the right panel of Fig.~\ref{fig:20101215-cs} shows only a snapshot of its cross-section.
Distribution of $B_Z$ component of magnetic field in the cross-section of the FRi3D model reveals slight asymmetry arising from the global 3D geometry of the structure.
The orientation of the invariant axis obtained via GS reconstruction differs from the local axis orientation of the FRi3D model by $17^\circ$.
The most obvious difference between the two magnetic field maps is their shape.
GS reconstruction produced an almost circular cross-section while FRi3D fit resulted in a strongly distorted pancake shape.
Another important difference is the estimated impact distance, \ie{}, the closest distance between the trajectory of the spacecraft and the axis of a CME.
FRi3D model fit produced the impact distance of $0.124$~AU while GS reconstruction estimated this parameter as $0.005$~AU.
By integrating the GS reconstructed magnetic field map using Eq.~\ref{eq:flux} we estimate the magnetic flux to be $3.3\times10^{12}$~Wb, which is significantly lower than $4.7\times10^{14}$~Wb predicted by the FRi3D model.
There are multiple possible explanations for such a mismatch.
On the one hand, since the shape of the flux-rope cross-section estimated by GS reconstruction does not take into account pancaking distortion its area is likely to be underestimated, which in turn could lead to underestimation of the total magnetic flux.
On the other hand, given the typical flux budget of an active region is of the order of $10^{14}$~Wb the FRi3D model seems to overestimate the magnetic flux released with a CME.
This issue in turn could result from underestimation of field lines twist near the edge of the structure and the usage of magnetic field distribution in the flux-rope cross-section described by Eq.~\ref{eq:lundquist-distribution}.

\begin{figure}[b]
\plotone{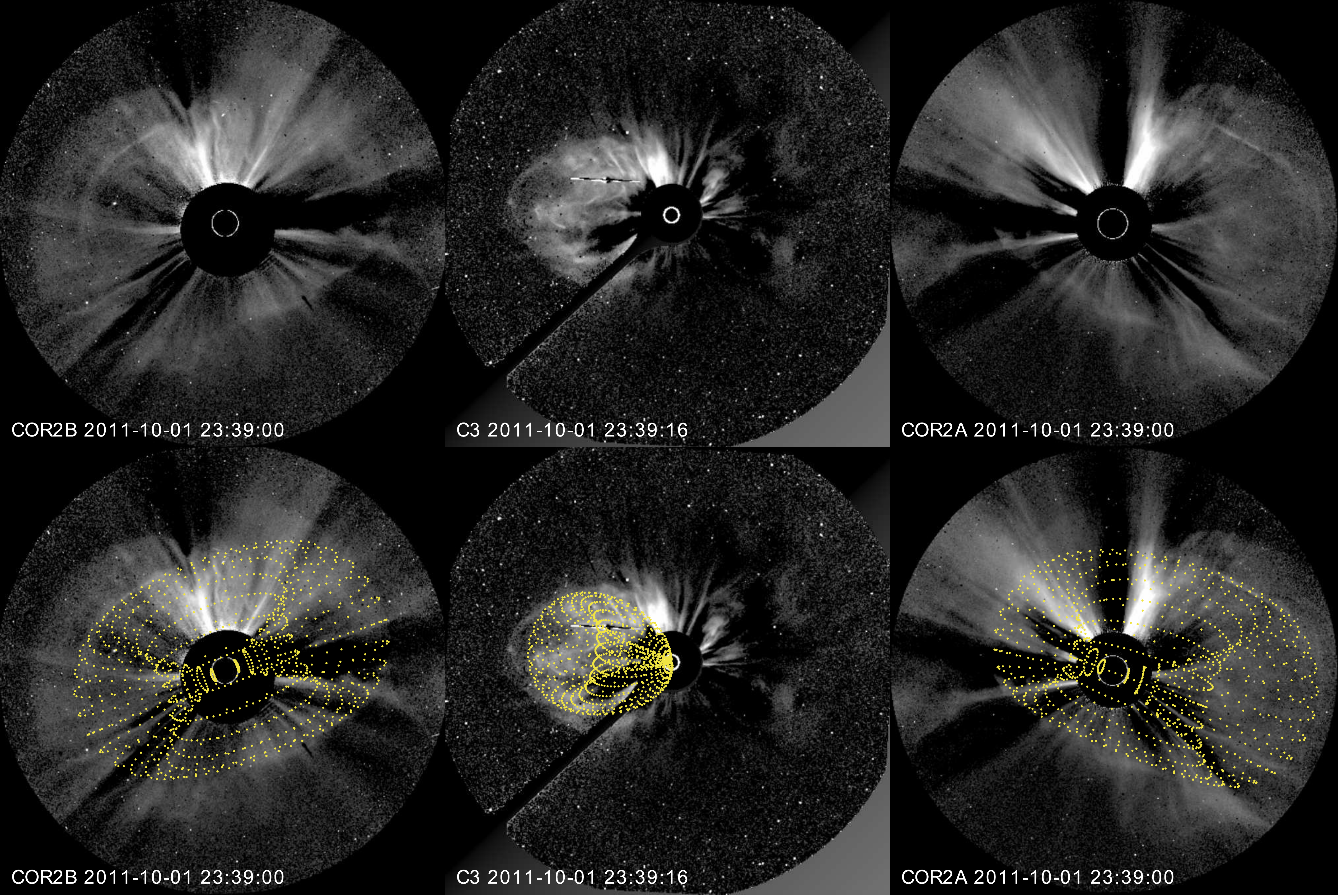}
\caption{Coronagraph images of the CME released on 1 October 2011. The images from left to right show observations from COR2 coronagraph of STEREO-B, C3 coronagraph of SOHO and COR2 coronagraph of STEREO-B respectively. Lower panels show the fitting of the FRi3D model to coronagraph observations.}
\label{fig:20111001-remote}
\end{figure}

In the second case study, we investigate a CME that was released on 1 October 2011 at 21:00~UT from the northern hemisphere.
This eruption is associated with a B9 class flare observed at $-117^\circ$ longitude and $19^\circ$ latitude in Stonyhurst coordinates \citep{Thompson2006}.
The event was observed by coronagraphs onboard SOHO and STEREO (see Fig.~\ref{fig:20111001-remote}).
It produced a full halo in STEREO-B field of view and a backside full halo in STEREO-A field of view.
This fast CME propagated through coronagraph field of view with projected speed of 1238~km/s and produced a clear shock wave front.

Lower panels of Fig.~\ref{fig:20111001-remote} show the fitting of the FRi3D model to coronagraph images from COR2 and C3 instruments.
Again, the model well describes the observations in a similar fashion as GCS.
Key geometrical parameters of the model fit are listed in Table~\ref{tbl:fit-params-201110}.

\begin{figure}[t]
\plotone{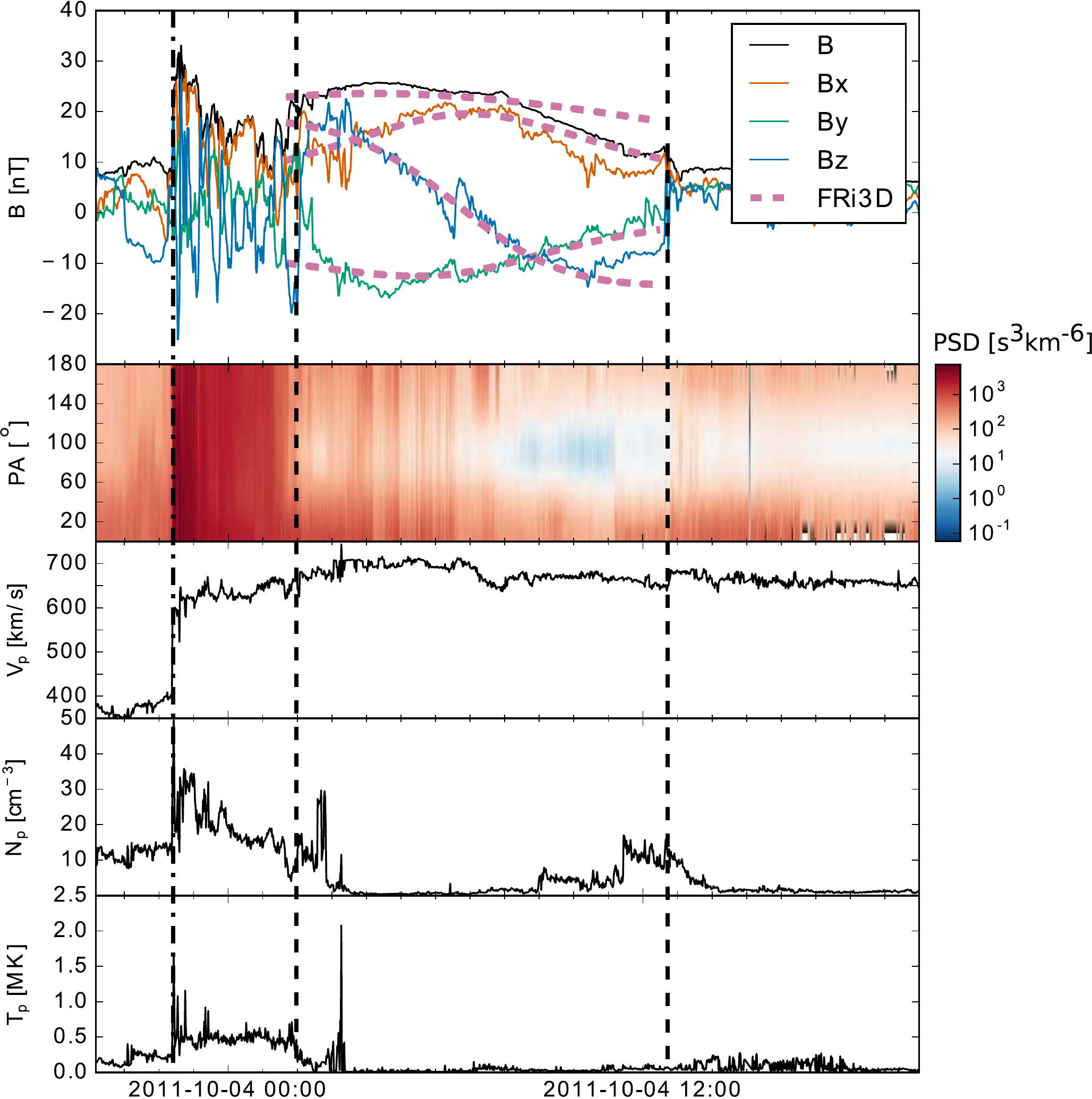}
\caption{\textit{In-situ} magnetic field and plasma measurements of the ICME launched on 1 October 2011 obtained by STEREO-B spacecraft. The panels from top to bottom show magnetic field, electron pitch angle distribution, plasma bulk speed, proton density and proton temperature. Magnetic field data are presented in HEEQ coordinates. Black vertical dash lines show the time range of magnetic obstacle. Purple dash curves show the FRi3D model fit.}
\label{fig:20111004-insitu}
\end{figure}

The interplanetary counterpart of this CME reached STEREO-B spacecraft on 3 October 2011.
The shock wave produced by the fast ejecta was registered on 3 October 2011 at 22:23~UT while the magnetic obstacle was measured between 02:00~UT and 12:40~UT of 4 October 2011 (see Fig.~\ref{fig:20111004-insitu}).

We carry the numerical fitting of the FRi3D model to \insitu{} data for this event using the same procedure as for the previous one with one exception.
Rapid decrease of total magnetic field in the rear part of the magnetic obstacle may be treated as a signature of flux-rope expansion.
Thus we take expansion into account by plugging linearly growing poloidal height $R_p$ defined by Eq.~\ref{eq:poloidal-height-profile-linear} into the model.

The results of the fit are shown in Fig.~\ref{fig:20111004-insitu} and the fitting parameters are listed in Table~\ref{tbl:fit-params-201110}.
The average euclidean distance between the modeled and real data is 3.53~nT.

\begin{deluxetable}{lrrrrrrr}
\tablecolumns{8}
\tablecaption{Parameters of the FRi3D model fits to remote and \insitu{} data for CME launched on 1 October 2011.\label{tbl:fit-params-201110}}
\tablehead{
\colhead{} & \colhead{$\theta$} & \colhead{$\varphi$} & \colhead{$\left\langle R_p \right\rangle/R_t$} & \colhead{$\varphi_{hw}$} & \colhead{$\gamma$} & \colhead{$n$} & \colhead{$\theta_p$}
}
\startdata
remote & 5.5 & -95.0 & 0.30 & 75.0 & 21.0 & 0.55 & 27.0\\
\hline
\insitu{} & -0.3 & -73.6 & 0.08 & 79.5 & 8.5 & 0.71 & 36.2\\
\nodata & \multicolumn{7}{l}{$\tau=1.2$, $\Phi=6.8\times10^{14}$~Wb, $V_{R_p}=36.7$~km/s}\\
\nodata & \multicolumn{7}{l}{East--West polarity, left-handed}
\enddata
\end{deluxetable}
According to our fits the CME deflected and rotated towards the solar equatorial plane and overexpanded in lateral and vertical directions.
The estimated speed of poloidal expansion of the structure is $V_{R_p}=36.7$~km/s, which seems a reasonable rate according to \insitu{} measurements.
The modeled CME has particularly low twist of 1.2 full rotations of magnetic field lines from footpoint to footpoint, which, however, is in good agreement with results reported by \citet{Hu2015}.
Our analysis showed that the CME experienced longitudinal deflection by $21.4^\circ$ eastwards.
However, a fast CME is expected to experience westward longitudinal deflection due to interaction with background magnetic field, which expands radially with slower solar wind and forms the Parker spiral \citep{Wang2004,Isavnin2013}.
Such a result has multiple possible explanations.
Firstly, error bars of the FRi3D model fits are not well-known yet.
Extensive statistical studies and comparison with MHD simulations, which are the subjects for follow-up research, would quantify the uncertainties of the fits. 
Secondly, estimated longitudinal deflection could be a result of non-radial expansion of the CME.
The difference between longitudinal location of the source region ($-117^\circ$) and direction of propagation estimated from remote observations ($-95^\circ$) shows that the CME is likely to have experienced eastward deflection by $22^\circ$ in the lower corona.
One could speculate that eastward drift of the structure slowly continued in the inner heliosphere.
Thirdly, interaction with the Parker spiral is more pronounced for weak magnetic field CMEs while the analyzed event exhibits relatively strong magnetic flux \citep{Kay2015}.

\begin{figure*}[t]
\plotone{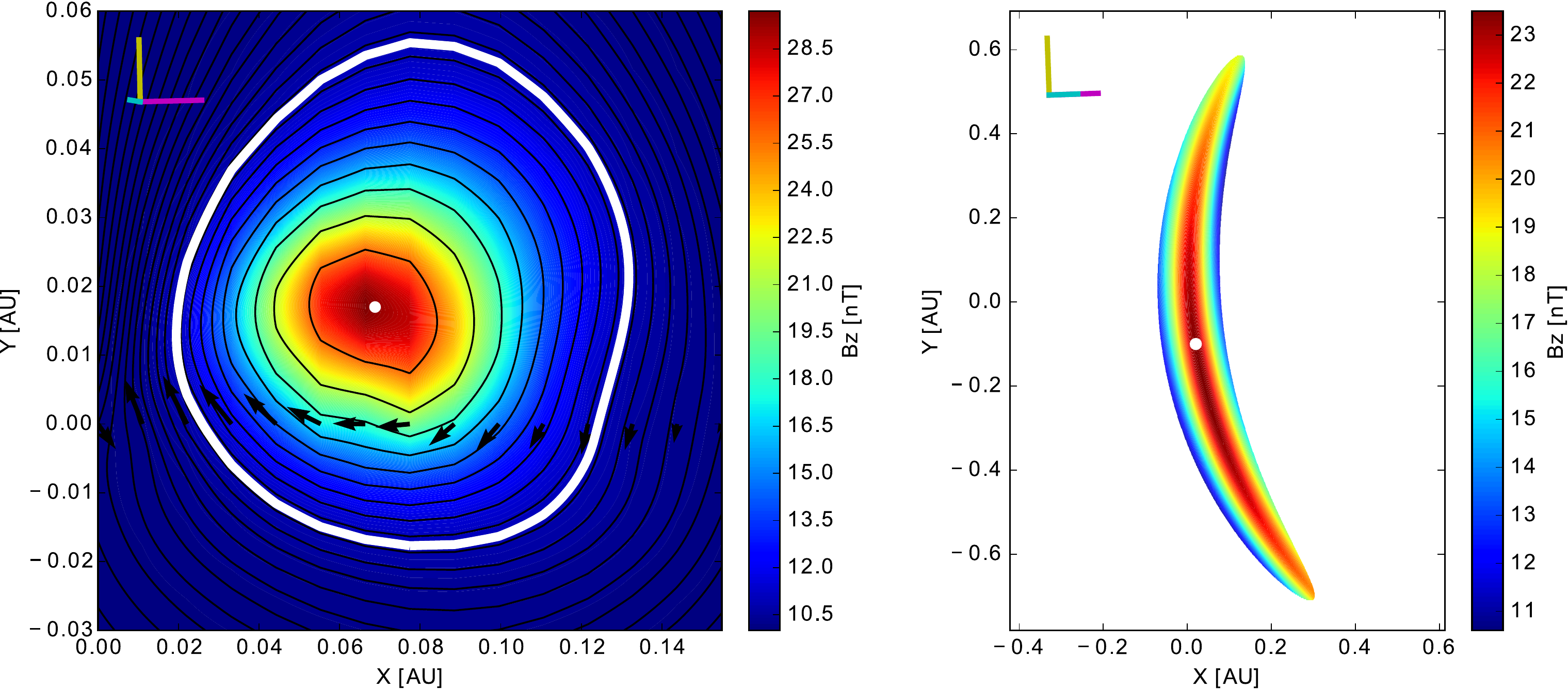}
\caption{Magnetic field maps obtained using Grad-Shafranov reconstruction (left) and FRi3D fitting (right) for the CME released on 1 October 2011. Magnetic field component parallel to the local axis orientation of the CME is color-coded. Local CME axis is marked with the white dot. Projected trajectory of the spacecraft goes along $Y=0$ line. The Sun is to the right. In the top left corners of the maps the projection of HEEQ coordinate system is shown as $X_{HEEQ}$ (cyan), $Y_{HEEQ}$ (magenta) and $Z_{HEEQ}$ (yellow). White solid curve in GS magnetic field map marks the boundary of an unperturbed part of the flux-rope. Black arrows show the projection of magnetic field measurements onto the cross-section plane.}
\label{fig:20111004-cs}
\end{figure*}

Fig.~\ref{fig:20111004-cs} shows side by side magnetic field maps calculated using GS reconstruction technique and FRi3D fitting respectively.
The orientation of the invariant axis obtained via GS reconstruction differs from the local axis orientation of the FRi3D model by $40^\circ$.
According to the FRi3D fitting the CME experienced strong expansion while GS reconstruction again resulted in almost circular shaped cross-section.
The impact distance according to the FRi3D model is $0.106$~AU, which significantly exceeds $0.017$~AU estimate by GS reconstruction.
By integrating the GS reconstructed magnetic field map using Eq.~\ref{eq:flux} we estimate the magnetic flux to be $2.7\times10^{12}$~Wb, again showing mismatch with $6.8\times10^{14}$~Wb predicted by the FRi3D model.
Possible reasons for this discrepancy are the same as we outlined earlier.

\section{Discussion and outlook}
We presented the first 3D model that is able to describe consistently both remote and \insitu{} observations of CMEs.
The FRi3D model encapsulates both global geometry and 3D magnetic structure of a CME and is able to reproduce its morphological and geometrical structure with high degree of complexity.
We applied the model for analysis of two example CMEs.
Independent model fits to remote and \insitu{} measurements of analyzed CMEs were found to provide consistent description of their global configuration.
The deduced properties of CME evolution were found to support earlier research on this subject, \eg{}, CMEs were found to deflect and rotate towards solar equatorial plane.

The FRi3D model uses relatively large amount of free parameters compared to traditional flux-rope fitting and reconstruction techniques which can lead to concerns about uniqueness of model fits.
However, after applying the model for analysis of two example CMEs we did not find that it is the case.
One possible reason could be the connection of the model to the Sun.
This geometrical feature poses a strong constraint on model parameters and is exempt from the majority of traditional local flux-rope fitting techniques.
Another plausible explanation is the clarity of events selected for case studies, \ie{}, ambiguities could rise for more distorted CMEs.

3D configuration of magnetic field with constant twist is constructed on the basis of empirical findings and thus it is not guaranteed or checked that the FRi3D model is force-free.
Multiple studies indicated that magnetic clouds associated with CMEs tend to have pressure gradients that cannot be explained with force-free approximation \citep{Mulligan2001,Hidalgo2002b,Moestl2009b}.
A possible contribution to these features could be the global geometry of CMEs, which is far from cylindrical.
Consequently, we do not treat the lack of force-free approximation as a disadvantage of the model.

Comparison with GS reconstruction showed that the FRi3D model does not seem to suffer from the typical shortcomings of conventional flux-rope fitting and reconstruction techniques, \ie{}, non-realistic shape of cross-section and underestimation of impact distance \citep{Riley2004}.
Nevertheless, distribution of the magnetic field from the edge to the center of the flux-rope cross-section, \ie{}, its minimum and maximum values, was found to be consistent with GS technique results.

The FRi3D model seems to overestimate the magnetic flux budget of a CME. 
This effect might result from underestimation of magnetic field lines twist near the edge of a flux-rope as well as inability of the Lundquist model (Eq.~\ref{eq:lundquist-distribution}) to properly describe the distribution of magnetic field in pancaked cross-sections.
In our further studies we will tackle this issue with at least the following approaches.
Firstly, we will test the version of the model with the constant twist rate, \ie{}, the amount of twist per unit length of a field line.
Secondly, we will check the possibility to plug in the Gold and Hoyle constant-twist nonlinear force-free model \citep{Gold1960} into the FRi3D.

3D modeling of CMEs is a relatively new area of space weather research and hence there are a lot of possible fitting strategies that can be applied to a model like FRi3D.
In fact, for testing purposes in our example CME studies we selected the worst case scenario, \ie{}, we fitted the model to remote and \insitu{} observations completely independently.
Such a strategy is good for demonstration of consistency of the model fits.
However, the full potential of 3D modeling is unleashed by fitting an evolving model to a series of CME observations.
In this scenario, a subset of free parameters of the model is assigned with evolution profiles expressed by any functions, \eg{}, linear evolution profiles represented by Eqs.~\ref{eq:toroidal-height-profile-linear} and \ref{eq:poloidal-height-profile-linear}.
The evolving model is then fitted to all available data, \ie{}, coronagraph images, heliospheric imager observations and \insitu{} measurements at any heliocentric distance.
Thereafter the fitted model can be used to predict further CME evolution.
Such an approach could be the first step to development of innovative space weather forecasting tools that would address prediction of both arrival time of a CME and magnetic field produced by it at a given point of interplanetary space.


\acknowledgments
The presented research was supported by the European Union Seventh Framework Programme (FP7/2007--2013) under grant agreement No.~606692 (HELCATS).

\bibliographystyle{aasjournal}
\bibliography{bibliography.bib}
\end{document}